\documentclass[%
 reprint,
 amsmath,amssymb,
 aps,
]{revtex4-2}

\usepackage{graphicx}
\usepackage{dcolumn}
\usepackage{bm}
\usepackage{braket}
\usepackage{amsmath}

\usepackage{natbib}
\setcitestyle{biber, phys, none}

\begin{document}

\preprint{APS/123-QED}

\title{Electron transport through a 1D chain of dopant-based quantum dots }

\author{Sumedh Vangara}
 \altaffiliation[Also at ]{Physical Measurement Lab, National Institute of Standards and Technology.}
\affiliation{%
 Poolesville High School
}%

\date{\today}

\begin{abstract}
Strongly interacting electron systems can provide insight into quantum many-body phenomena, such as Mott insulating behavior and spin liquidity, facilitating semiconductor optimization. The Fermi-Hubbard model is the prototypical model used to study such systems. Recent research, however, has shown that the extended Fermi-Hubbard model, which accounts for long-range interactions, is more accurate, especially for systems far from half-filling. In this study, we use the extended Fermi-Hubbard model to mathematically analyze charge transport through a lattice of quantum dots. One-dimensional chains with spinless electrons and source-drain bias are observed, focusing on the transition between the ground state and the first excited state. Level repulsion decreases the expected energy levels of anticrossings as the hopping onto the chain tends to the hopping within the chain. The distribution of charge density along the chain is characterized in terms of the hopping, nuclear, and Coulomb parameters and novel plasmonic behavior is analyzed. Minor perturbations in electron transport are identified, corresponding to the one-dimensional nature of the observed systems. This research will lead to a better understanding of electron behavior in silicon-doped semiconductors, like the formation of correlation-induced band gaps, and open the door to using the extended Fermi-Hubbard model as a more accurate alternative to study quantum many-body systems.

\end{abstract}

\maketitle

\section{\label{sec:level1} Introduction}

Strongly interacting systems are currently the foremost area of research in condensed matter physics. Simulating strongly interacting fermions on a lattice is central to understanding quantum many-body phenomena such as high-temperature superconductivity \cite{high-temperature-superconductivity}, Mott insulators \cite{mott-insulator}, and topological phase transitions \cite{topo1, topo2, topo3}. Strongly interacting simulations also may provide further insight into improving silicon nanoscale transistors, in particular for quantum computing as it could allow for low-power dissipation and high packing density \cite{transistor}. 

Here, we consider atomic-scale solid state systems, as a first principles theory would require the treatment of thousands of interacting electrons and and strong confinement. Such systems include metal atoms on a surface or in a break junction \cite{metal1,metal2,metal3}, chains of dopant atoms in silicon \cite{silicon1,silicon2,silicon3}, quantum dots or "artificial" atoms \cite{dot1,dot2}, or dangling bonds \cite{dangling1,dangling2,dangling3}. Recent studies have also created analog quantum simulators to circumvent simulating complex quantum systems intractable with classical computers \cite{silicon1,silicon2,silicon3,silicon4}

The Fermi-Hubbard model is the prototypical example of a system with strongly interacting electrons \cite{hubbard1,hubbard2,hubbard3}. In this paper, it will rely on three parameters: the hopping parameter $t$, Coulomb parameter $c$, and nuclear parameter $n$. The Fermi-Hubbard model and its variants are believed to describe a wide range of phenomena including unconventional superconductivity \cite{unconventionalsuper}, quantum spin liquids \cite{spinliquid}, and Nagaoka ferromagnetism \cite{Nferromagnetism}. However, due to the lack of numerical solutions for multi-electron systems, these phenomena are not fully understood, which has led to recent literature in which the Fermi-Hubbard model has been used for systems such as dopants in silicon lattices. Most previous work on the physics of dopant-based systems rely on assumption that the Fermi-Hubbard model is satisfactory if the Hubbard Hamiltonian only contains on-site interactions. However, recent literature \cite{silicon3} has shown that models with only the on-site interaction are not sufficient when the system is far from half-filling. In addition, recently, an extension of the Fermi-Hubbard model that includes off-site interactions, the extended Fermi-Hubbard model, has gained traction through its use in contexts such as ultracold fermions in optical lattices \cite{coldfermion1,coldfermion2,coldfermion3} and transition metal dichalcogenides \cite{transitionmetaldi}. It has also gained popularity due to a recent breakthrough in which an analog quantum simulator for the two-dimensional extended Fermi-Hubbard model was developed using dopant-based quantum dots \cite{silicon4}.

In this paper, we observe atom-scale systems that are metallic and can support many-body systems. We mathematically model, using the extended Fermi-Hubbard model, small metal-like atom scale systems. In particular, we model one-dimensional chains of quantum dots, which represent dopant phosphorus atoms in a silicon lattice. We also allow for source-drain bias, allowing electrons to leave and enter the chain. But we only consider spinless electrons, electrons of one spin, so Pauli exclusion ensures at most one electron can exist per site. All diagrams and references to measures of charge are done with units of the elementary charge $e$. We will refer to an $n$-site chain as a source connected to $n$ sites in a line connected to a drain, so the entire system has $n+2$ $\emph{total}$ sites. To simplify our system, we only look at transitions between the ground and first excited states and observe wavefunctions where there is an electron in the source that is moving into the chain. This ensures that electron transport will always arise through these chains of quantum dots as an electron will always be transitioning from the source site to the first site in the chain.

\section{Theory}
\subsection{Hamiltonian}

The Fermi-Hubbard model is employed with both on-site and off-site interactions: $$\sum_{\langle i,j \rangle_{nn}}{t \left( c_i^{\dag}c_j + c_ic_j^{\dag} \right)} + \sum_{i}{V_{nuc}(i)n_i} + \sum_{\langle i,j \rangle}{V_{ee}^{D}(i,j)n_in_j}$$
The first summation corresponds to the kinetic term. The hopping parameter is $t$. The summation goes over all adjacent pairs of sites and $c_i^{\dag}$ is the creation operator for an electron at site $i$ while $c_i$ is the corresponding destruction operator. The second term in the summation governs the potential due to interactions between the electrons and lattice sites, with $n_i=c_i^{\dag}c_i$ being the electron number at site $i$. The third sum covers the interaction between all pairs of electrons on the lattice. Since all electrons are identical, there is no on-site interaction between electrons. In addition, we assume that the on-site energy of all the sites that do not have bias are equal, specifically $0$ for convenience. 

Interactions are modeled using the long-range Coulomb interaction. The interaction energy $V_{nuc}$ between an electron at site $i$ and the atoms is: $$V_{nuc}(i) = - \sum_{j}{\frac{\lambda_{nuc}Z}{|i-j|+ \xi_{nuc}}}$$
The long-range electron-electron interaction energy $V_{ee}^{D}$ between electrons at sites $i,j$ is defined as: $$V_{ee}^{D}(i,j) = \frac{\lambda_{ee}}{|i-j|+ \xi_{ee}}$$ $\lambda_{nuc}$ and $\lambda_{ee}$ are scale factors that account for the length scale for nearest-neighbor separation and any dielectric screening. $\xi_{nuc}$ and $\xi_{ee}$ are cutoffs that include the spread of the electron orbital.

The direct Coulomb interaction, $V_{ee}^{D}(i,j)$, between electrons on sites $i,j$ couples charge densities $n_i,n_j$. Scattering effects could also be included where one electron scatters in the Coulomb potential of another electron and hops between sites. This would create another term in the Hubbard Hamiltonian based on Coulomb-induced hopping, depending on local electron density. This hopping however is weak for anything other than nearest-neighbor hopping. The induced, nearest-neighbor hopping is also weaker than the interaction between electrons at two neighboring sites, so we ignore all Coulomb induced-hopping. Scaterring effects when both electrons hop are also ignored. 

Exchange effects are also included. When a pair of electrons on adjacent sites switch sites due to the Coulomb interaction, we include exchange by reducing the strength of the Coulomb interaction: $$V_{ee}^{exch} \left(i,i_{nn} \right) = - \lambda_{exch} V_{ee}^{D} \left( i,i_{nn} \right)$$ where $\lambda_{exch}$ is the scale of the nearest neighbor exchange and $i_{nn}$ is a site neighboring site $i$.

\begin{figure}[htp]
\includegraphics[width=8.6cm]{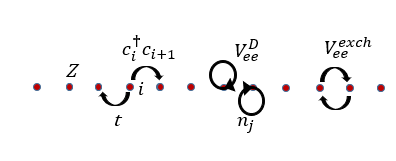}
\caption{Diagram of the model. $Z$ is the nuclear charge at each site, $t$ is the hopping between nearest neighbor sites, $V_{ee}^{D}$ is the electron-electron interaction, and $V_{ee}^{exch}$ is the exchange interaction. \cite{diagram}}
\end{figure}

Figure 1 shows the geometry of the system. The dots represent quantum dots, which are the sites, and electrons can move between these sites. In particular, the system depicted in Figure 1 has $12$ sites excluding the source and drain. For a system with $N$ sites and $m$ electrons, the number of possible states is $\binom{N+2}{m}$. Thus, for every four sites that are added, the number of possible states increases by an order of 1 or 2 magnitudes. The storage of more than a few hundred lowest energy eigenstates can become expensive. For instance, a system with $20$ sites that is half filled has nearly $200,000$ total states. However, for shorter chains, such as half-filled ones with $14$ sites, the entire Hamiltonian can be diagonalized and all the eigenstates can be calculated. Adding other effects such as spin or coupling to a parallel chain to model two-dimensional effects are straightforward, but since they would reduce the possible search space and are irrelevant to observing these specific systems, we choose to ignore them. We also use the values of $\xi_{ee}=\xi_{nuc}=0.5$, which correspond to the spread in the electron orbital with half lattice spacing. This has little effect on the nature of the system. With distances of $2-5$ nanometers, the nearest neighbor Coulomb interaction is usually between $10$ and $100$ meV while hopping is usually between $1$ and $10$ meV.

\subsection{Time-Dependence}

To carry out time-dependence calculations using the Hamiltonian, we use Schrodinger's equation. In what follows, we will show how to construct the Hamiltonian for simple systems, in particular systems with no charge interactions.  

The matrix for the Hamiltonian, $\hat{H}$, describes transitions between two states. With $n$ sites and $k$ electrons, it will contain $\binom{n+2}{k}$ rows and columns. Each row and each column correspond to a state, specifically a configuration of electrons on a set of sites in the chain. The correspondence can be chosen arbitrarily, but in this paper, we will choose them lexicographically, according to the numbering of the sites, with the only exception being that the source will come before all the sites on the chain. Regardless of the correspondence chosen, the $i$th column and $i$th row should also correspond to the same state of the system. 

The matrix is then filled out as follows: for terms on the main diagonal, the energy of the state should be placed in it. For all other terms $\hat{H}_{ij}$, if there is a transition from state $i$ to state $j$, that matrix element should be the hopping value $t$. Otherwise, it should be $0$. For example, for a chain with $1$ site, along with a source and drain, has Hamiltonian \[ \hat{H} = 
\begin{bmatrix}
V_0 & t & 0\\
t & 0 & t\\
0 & t & V_0
\end{bmatrix}\]
where the main diagonal has a $V_0$ term on the top left and bottom right cells corresponding to source-drain bias and the two diagonals neighboring the main diagonal all have the term $t$. These values of $t$ correspond to an electron hopping between adjacent sites and thus the system changing states. In general, for systems with one electron, we have \[\hat{H} = \begin{bmatrix}

V_0 & t_1 & 0 & 0 & 0& \ddots & 0\\
t_1 & 0 & t & 0 & 0& \ddots & 0 \\
0 & t & 0 & t & 0 & \ddots & 0 \\
0 & 0 & t & 0 & t & \ddots & 0 \\
0 & 0 & 0 & t & 0 & \ddots & 0\\
\vdots & \ddots & \ddots & \ddots & \ddots & \ddots & \vdots\\
0  &  0 & 0 & 0 &  \dots & t_1 & V_0
\end{bmatrix}
\]
Note that since the hopping on and off the chain can be different than the hopping in the chain, we have designated $t$ as the hopping within the chain and $t_1$ as the hopping on and off the chain in $\hat{H}$. 

As the number of electrons increase until the system becomes half-filled, the number of cells that have the value $t$ also increase. This is because there are more states possible in the system as $\binom{n}{k}$ is increasing while $k < \frac{n}{2}$. Moreover, more electrons implies more states a state $i$ can transition to.

$\hat{H}$ is then diagonalized as finding the eigenvectors of $\hat{H}$ provides all the possible time-independent wavefunctions $\psi$. We first find $\lambda$ so $\det{(\hat{H} - \lambda I)} = 0$. To find this determinant, we can leverage the specific structure of $\hat{H}$ detailed before to construct a faster method than standard methods like Laplace expansion. In particular, the principal diagonal has only two non-zero terms: the first and last, corresponding to the source-drain bias applied in the system. The upper and lower diagonals alternatively have only non-zero elements. These elements are almost all the same except for the first and last elements in each of these diagonals, as they correspond to the hopping on and off the chain which may be different than the hopping in the chain. Overall, $\hat{H}$ is a tridiagonal matrix, which allows for calculations to be computationally more efficient. 

To find the determinant, we can leverage the tridiagonal nature of $\hat{H}$ along with results from \cite{determinant} to set up a recursive relation where $a_k$ is the $k$th term on the major diagonal:
\[f_k(\lambda) = (a_k - \lambda)
f_{k-1}(\lambda) - t^2f_{k-2}(\lambda)\]
which then allows us to calculate $f_m$ as the characteristic polynomial of $\hat{H}$ where the dimension of $\hat{H}$ is $m$. From here, we can use a root-finding method on our characteristic polynomial such as Newton's method or the Jenkins-Traub algorithm. We use the Durand–Kerner method. 

We need to then solve the general equation of \[\hat{H}a_i = \lambda_ia_i \implies (\hat{H}-\lambda I)a_i = 0\] for $a_i$. To solve this, we use the general method for solving $Ax=b$. In particular, we find the Moore-Penrose inverse $A^+$ and the associated nullspace projector \[P = I - A^+A\] With these matrices, the general solution can be written as \[x = A^+ b + Py\] where the vector $y$ is arbitrary. Substituting $A=\hat{H}-\lambda I, x=a_i, b=0$ finishes. 

To time-evolve the system, we compute \[e^{i\hat{H}t} \ket{S_x}\] where $S_x$ corresponds to the $x$th site. Finding the charge density at site $x$ easily follows with \[\bra{S_x}e^{i\hat{H}t} \ket{S_x}\] To calculate $e^{i\hat{H}t}$, the Taylor series is used: \[e^{A} = \sum_{n=0}^{\infty} \frac{A^n}{n!}\] The diagonalization of $\hat{H}$ is relevant here as
if $A = PDP^{-1}$, then the Taylor series turns into \[e^A = \sum_{n=0}^{\infty} \frac{PD^nP^{-1}}{n!}\]

All code to carry out computations was written in C. They are all either built-in libraries or programs we created for the purpose of calculating characteristics about these systems.

\section{Results and Discussion}
\subsection{Energy Levels}

Level repulsion skews the effects of the expected energy levels for anticrossings. The expected energy levels for crossings are linear in each of the hopping, nuclear, and Coulomb parameters, and follows this throughout all systems.  
\begin{figure}[h!]
\includegraphics[width=8.6cm]{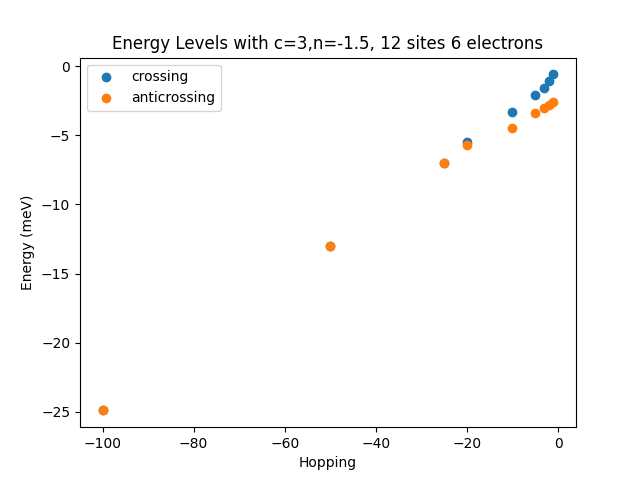}
\caption{Systems with $12$ sites, $6$ electrons, $c=3$, and $n=-1.5$ are depicted for various hopping values. In particular, the energy levels for their crossings and anticrossings, which occur when the source and drain are uncoupled and coupled respectively with the chain, are graphed. The crossings and anticrossings initially follow the same linear path, but as the hopping in the chain, $t$, tends to the hopping on and off the chain, $-1$, the energy levels for anticrossings plateau. All observed systems follow this trend. }
\end{figure}

Figure 2 demonstrates one such system where varying the hopping leads to level repulsion as the hopping in the chain tends to the hopping on and off the chain. The coupling between the ground energy state and first excited energy state is split apart by hopping, proportional to the value of the hopping parameter. 
This occurs the closer the hopping in the chain, $t$, tends to the hopping on and off the chain, $-1$, and causes the energy levels of the anticrossings to decrease below the energy level for the corresponding crossing. 
The hopping where level repulsion begins to have a noticeable effect is also proportional to the nuclear and Coulomb charges: when the forces are larger, the crossing and anticrossing energy levels deviate at hopping values that are more distant from the hopping onto and off the chain, in this case $-1$. For example, when the Coulomb and nuclear values are $0$, the crossing and anticrossing energy levels deviate when the hopping onto the chain equals the hopping in the chain, while Figure 2 shows that the energies deviate when the hopping within the chain approaches $-10$. 

These results are confirmed by previous research. In particular, it has been shown that level repulsion decreases the expected energy levels for anticrossings. This deviation is also more pronounced when the nuclear and Coulomb charges are larger, and the hopping on and off the chain doesn't need to be as close to the hopping in the chain to have the same effect \cite{diagram}.

\subsection{General Characteristics} 

In what follows, we will observe systems with $12$ sites and $3$ electrons. We will again refer to site $13$ as the source, site $14$ as the drain, and sites $1$ through $12$ as the sites on the chain from left to right. 
\begin{figure}[h!] 
\includegraphics[width=8.6cm]{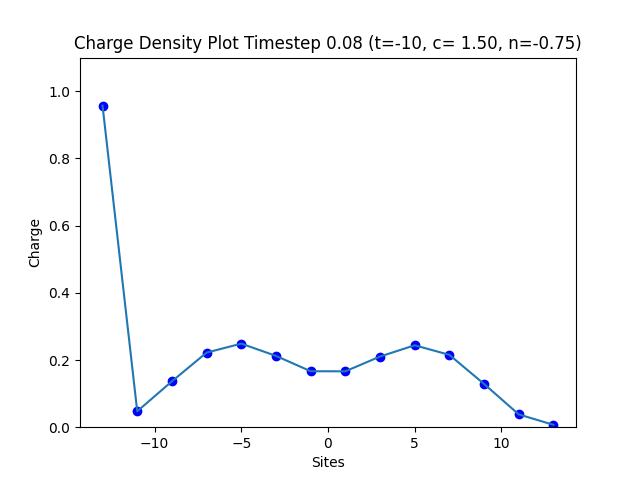}
\caption{Shows the distribution of electrons at an arbitrary timestep for an arbitrary system. The lines connecting points do not represent anything, but are used to visualize the overall trends in charge distribution. There are two local maxima in the chain around the $5$th and $10$th sites. There is a dip in charge density in the middle of the system. The charge in the source is elevated because the system was chosen to start with a whole electron in the source. }
\end{figure}

\begin{figure}[h!] 
\includegraphics[width=8.6cm]{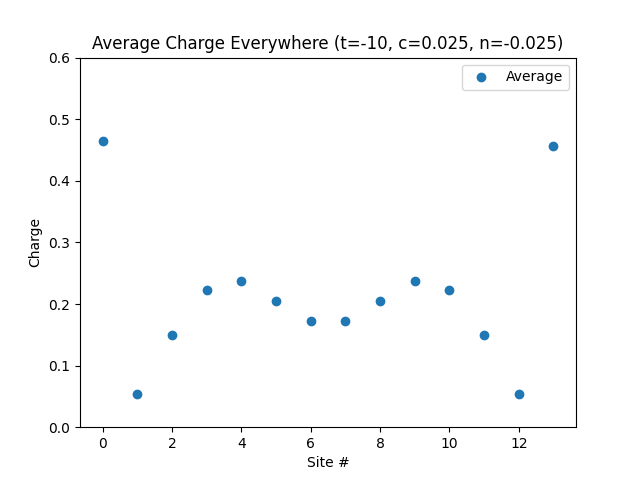}
\caption{ 
Depicts the average charge across all timesteps for each site on the chain for an arbitrary system. In particular, there are two relative extrema in the chain corresponding to the $3$ electrons in the chain, while the source and drain have elevated charge values. 
}
\end{figure}

The general structure of the electrons in the chain are depicted in Figure 3. Over time, the source and drain vary frequently, which we will delve into later in this paper. Regarding the other $12$ sites, a strong pattern emerges. There are two bulges, centered about one-third and two-thirds of the way into the chain, with a depression about halfway into the chain. Charge density also tapers off at the $1$st and $12$th sites. Charge density in the source and drain effectively vary independently. 

This structure can also be seen when observing the average charge at each site across the allotted time, shown in Figure 4. The average distribution of charge across the chain is symmetric about the center of the chain, and the average charge in the source and drain are about $0.5$. 

This structure occurs because of the nuclear charges in the system along with the fact that there are precisely $3$ electrons in the system. The nuclear charges pull electrons into the chain as electrons at the center of the chain will feel the strongest force from the nuclear charges while electrons at the ends of the chain feel the weakest force. However, in these systems, the electron-electron interaction is strong enough to prevent all the electrons from filling the middle of the chain. This causes the optimal structure to center about two points, symmetric about the chain. Systems with weaker electron-electron interaction should see the electrons center about the center of the chain, and systems with more electrons should see the electrons center about more than two evenly spaced points on the chain. 

\begin{figure}[h]
\includegraphics[width=8.6cm]{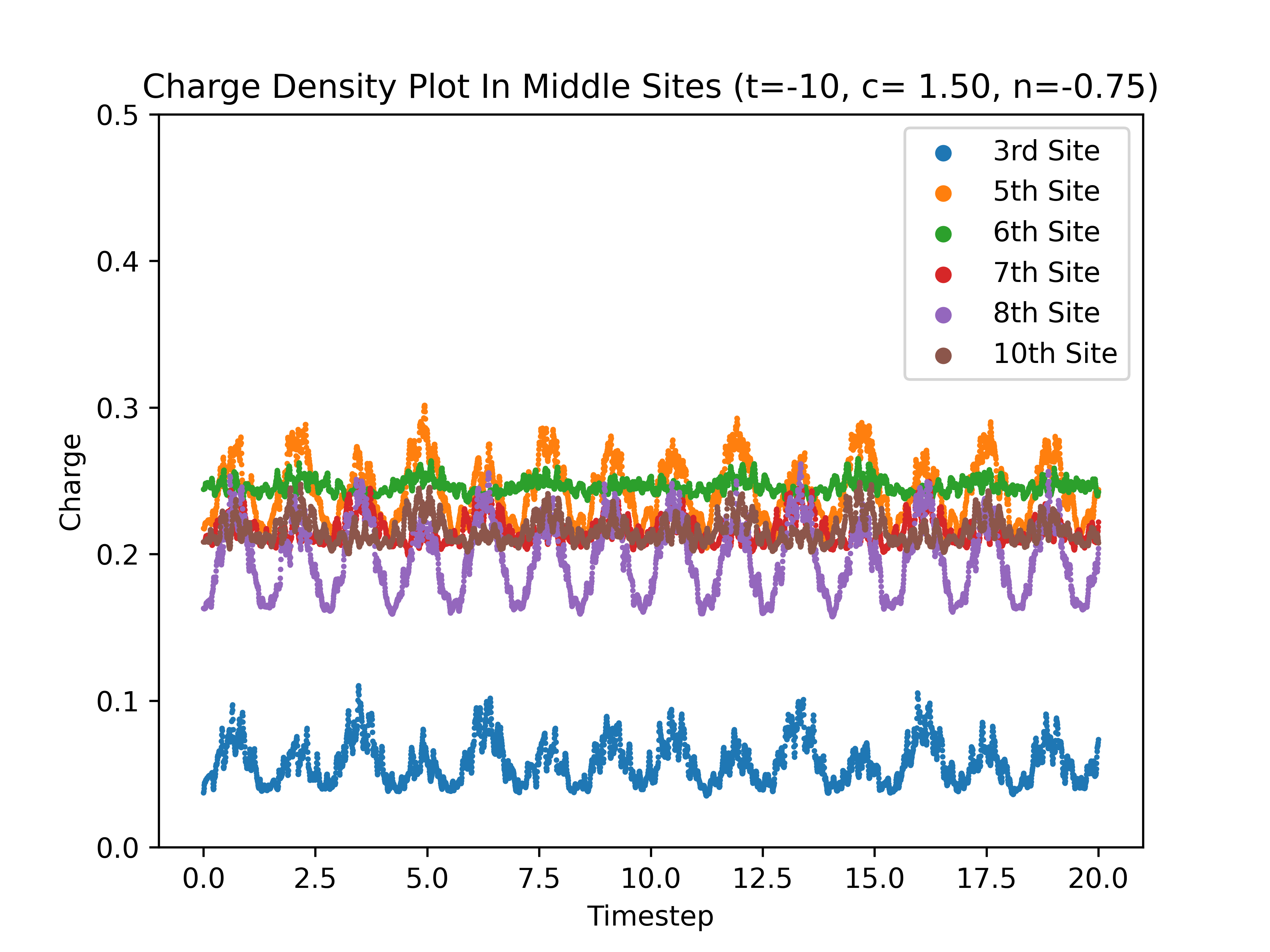}
\caption{ Graphs the charge density in several sites for an arbitrary system. Demonstrates that the charge density on a site varies the further away the site is from the center of the chain. }
\end{figure}

The charge densities at sites in the chain also vary over time. Figure 5 demonstrates how the location of a site in the chain affects the variance, or amplitude, of the charge density oscillation at that site. The sites that are further away, such as the $10$th site or $3$rd site, have a much larger variance in charge density than the sites closer to the center of the chain such as the $6$th and $7$th sites. Combined with the charge density graphs being sinusoidal, this suggests the electrons all oscillate on the chain as a collective. In the system in Figure 5, only the center sites having minimal variance in charge density suggest that the mass of electrons occupy about half of the chain at a time. Increasing the number of electrons would cause more sites to have charge densities that vary minimally, and decreasing the number of electrons would cause more sites to have charge densities that vary even more. 

Now we will observe the sinusoidal nature of these systems. All systems resembled, in some form, a sinusoidal motion with minor or major perturbations. 

\begin{figure}[h]
\includegraphics[width=8.6cm]{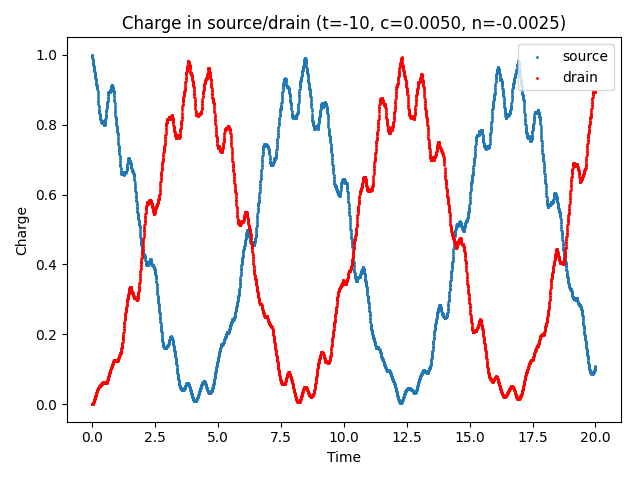}
\caption{ The charge in the source and drain when $t=-10,c=0.005,n=-0.0025$. The system is roughly sinusoidal with minor perturbations.  }
\end{figure}

Figure 6 shows one such periodic system, or plasmonic system, with minor perturbations. The overall system is periodic because of the oscillation of all the electrons moving across the chain. This creates the oscillation shown in Figure 6 as the ends of the plasmon periodically fall into the source and drain. This is also supported by observing the charge in the source and drain combined, shown in Figure 7. 

\begin{figure}[h!]
\includegraphics[width=8.6cm]{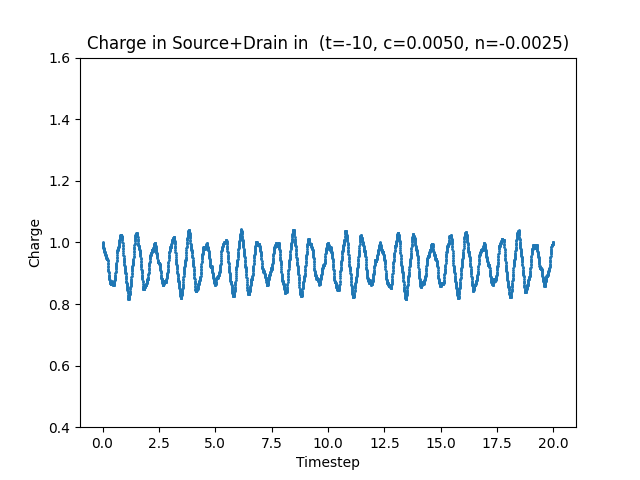}
\caption{The sum of the charge in the source and in the drain over time for the same system from Figure 5. The graph is sinusoidal. }
\end{figure}

The periodic nature of the total charge in the source and drain is found across all observed systems and is in line with previous observations on systems similar to these \cite{diagram}. 

\begin{figure}[h!] 
\includegraphics[width=8.6cm]{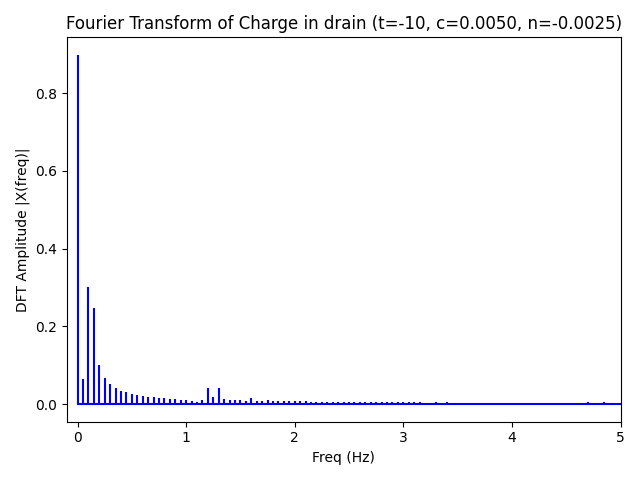}
\caption{A Discrete Fourier Transform (DFT) of the graph of the charge in the drain when $t=-10, c=0.005,n=-0.0025$. There is one spike corresponding to the plasmon, or main oscillation that can be seen in Figure 5, and smaller peaks that correspond to the imperfections in the main oscillation due to perturbations.}
\end{figure}

The minor perturbations on the other hand occur because of the geometry of the chain. In two-dimensional systems for instance, these perturbations do not occur as the second dimension allows for electrons to move around other electrons, but this is not true for a one-dimensional chain. For instance, if one electron was on site $1$ and another on site $3$ and the electron on site $1$ moves to site $2$, the electron on site $3$ may push the electron on site $2$ back to site $1$. These minor perturbations create the jagged structure of the charge curves in Figure 6, and can also be seen by taking a Discrete Fourier Transform of Figure 6, which is shown in Figure 8. 

All systems follow this same Discrete Fourier Transform pattern with one spike besides $0$ and none or some other frequencies with much smaller amplitude, and therefore impact on oscillation. The large spike corresponds to the overall oscillation while the frequencies with smaller amplitudes correspond to minor perturbations as mentioned before. 
\begin{figure}[h]
\includegraphics[width=8.6cm]{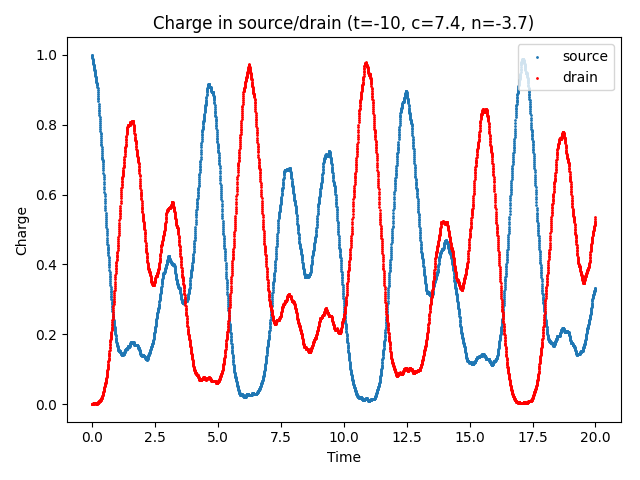}
\caption{Shows the charge in the source and drain over time in a system with $t=-10,c=7.4,n=-3.7$. In particular, while systems like these are periodic, they do not appear purely sinusoidal. }
\end{figure}

While all systems are periodic, some greatly deviate from purely sinusoidal, one of which is shown in Figure 9. This occurs because this system dephases immediately, while systems like the one in Figure 6 do not dephase immediately and therefore remain purely sinusoidal except for minor perturbations. The contrast of these two systems demonstrate that there are two different regimes: one where dephasement happens immediately in the start and one where dephasement does not happen at the start. Each of these regimes exist in continuous ranges of $c,n$. For instance, for an arbitrary hypothetical system, dephasement could occur immediately between $c=2,n=1$ to $c=4,n=2$ but not immediately from $c=4,n=2$ to $c=6,n=3$. In other words, there do not exist discrete, non-continuous points where systems change from dephasing to not dephasing or vice versa. Thus, choosing the proper parameters of $t,c,n$ can determine the speed at which a system will dephase, and therefore deviate from mostly sinusoidal motion. Future research should be conducted on what ranges constitute immediate dephasement and non-immediate dephasement, and how sharp the edges of these ranges truly are.

\subsection{Transport in Source and Drain}

The total charge in the source and drain changes as the Coulomb and nuclear charges change. Specifically, with small values of $c$ and $n$, the oscillation has a large amplitude and its midline is above $1.0$. Gradually, as the magnitudes of $c,n$ increase, the midline becomes $1.0$ while the amplitude decreases. After a certain threshold, any increase in the magnitudes of $c,n$ causes the midline to decrease below $1.0$ while the amplitude again becomes larger as shown in Figure 10. 

\begin{figure}[h]
\includegraphics[width=8.6cm]{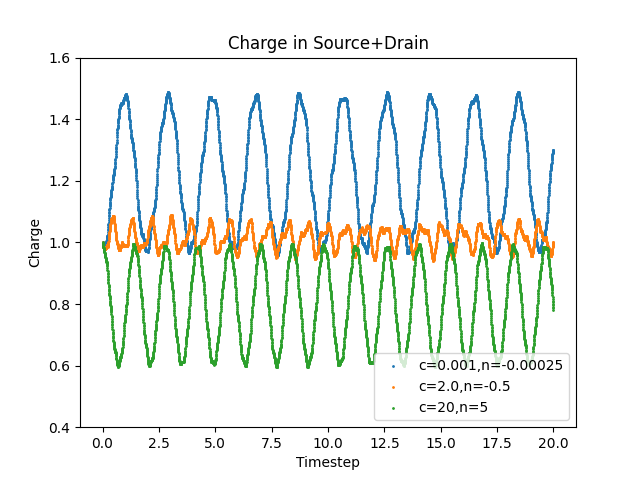}
\caption{Shows the graph of the total charge in the source and drain over time. It shows it for three different systems: $c=0.001,n=0.00025$, $c=2.0, n=0.5$, and $c=20.0, n=5.0$. All oscillate with different midlines and periods. Note that there is a net-positive charge in the system. }
\end{figure}

Thus, there exists a continuous set of values for $c,n$ that minimize the variance of total charge in the source and drain. This phenomenon occurs because when the nuclear and Coulomb charges are small, the site charges are not strong enough, in magnitude, to retain electrons. This allows the plasmon to spread out so the net charge in the source and drain is over $1.0$ on average, and the minimal forces allow for more movement creating a large amplitude. As the forces become stronger, the plasmon is constricted. The electrons are not as free as before because of the stronger forces, so this brings down the midline as the electrons cannot cover as much of the chain. Simultaneously, the amplitude of oscillation of the total charge in the source and drain is minimized as the distance the plasmon oscillates perfectly aligns with the length of the chain, minimizing the amplitude when the midline is centered at $1.0$. In other words, one part of the plasmon is entering the source while the other is close to leaving the drain and vice versa at this intermediary stage. When the nuclear and Coulomb charges are further increased, this further restricts the plasmon so that it is even less distributed across the chain. Thus, when it is at one end of the chain, the charge in the source and drain will be approximately $1.0$, but when the plasmon is in the chain, the charge in the source and drain will be much less. This correspondingly reduces the midline as the average charge in the source and drain will be less than both $1$ and the average charge in the source and drain for the other two types of systems mentioned before. This also increases the amplitude again as the minimum charge in the source and drain hits a number far less than $1$ while its maximum is about $1$. 

\begin{figure}[h!]
\includegraphics[width=8.6cm]{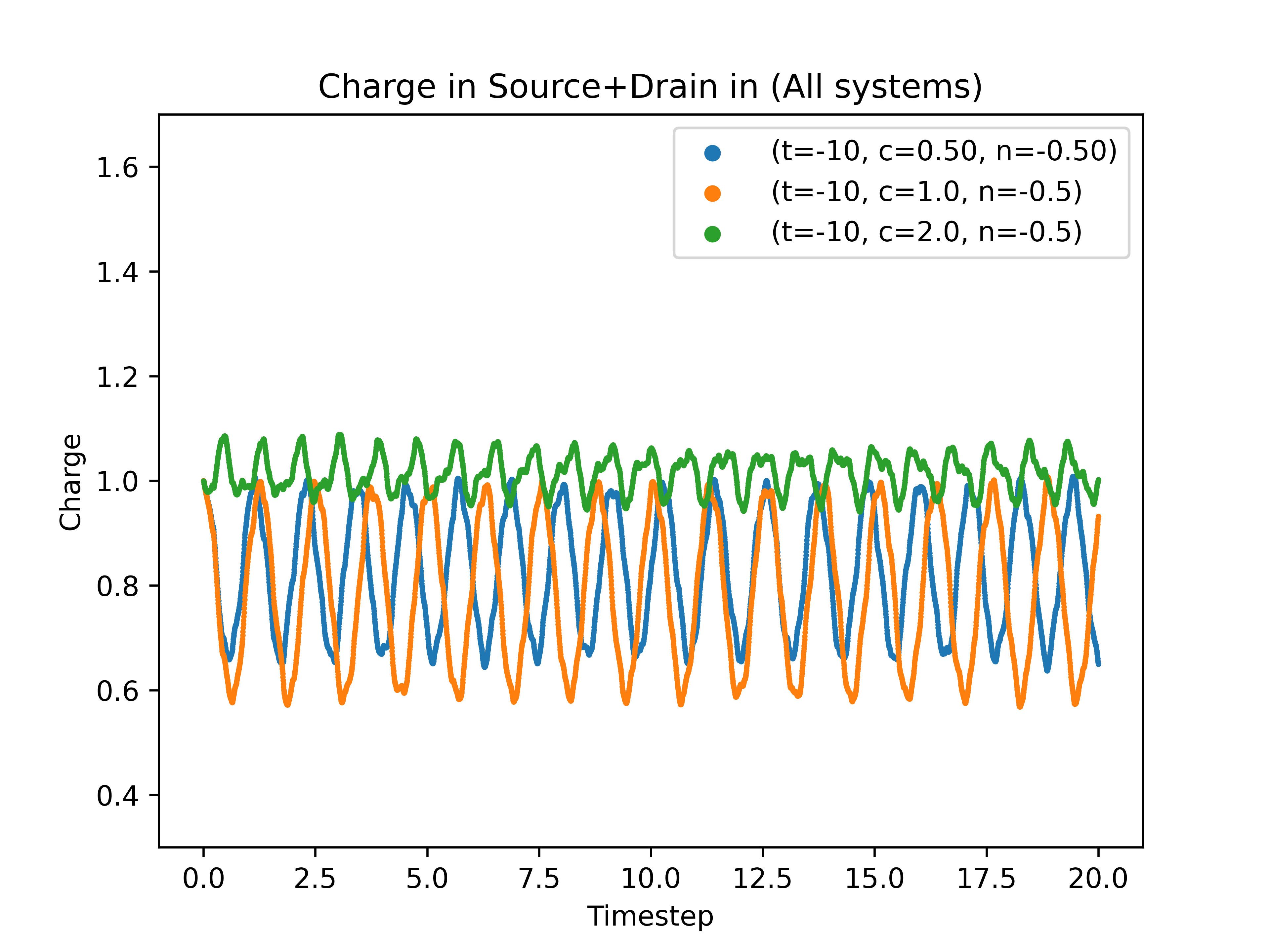}
\caption{The sum of the charge in the source and in the drain over time for three different systems. When the ratio between the Coulomb charge and nuclear charge is $-4$, or when the system is net neutral, the amplitude is much smaller and the midline is much higher than when the ratio between the Coulomb charge and nuclear charge is $-2$ or $-1$. In particular, when the ratio is $-2$ or $-1$, the graphs are very similar. These two systems are net positive. }
\end{figure}

When observing the total charge in the source and drain while fixing the nuclear charge and varying the Coulomb charge, a different pattern emerges. 

We will denote $F = \frac{c}{n}$. It is important to note that as $\frac{12}{3} = 4$, these systems with $12$ sites in the chain and $3$ electrons are net neutral when $F = -4.0$. Thus, when $F = -1.0$ or $-2.0$, the system is net positive.

The phenomenon in Figure 11 occurs precisely because the system is net neutral when $F= -4.0$. At net neutrality, since the chain itself is net neutral, electrons on the source and drain are less likely to hop into the chain than if the system, and thus chain, was net positive, such as when $F$ is $-2.0$ or $-1.0$. This is also supported by the minuscule amplitude for the graph when $F=-4.0$, as there is less electron movement occurring compared to when $F=-1.0$ or $F=-2.0$. Thus, net neutral systems result in far less pronounced oscillations. 

\subsection{Varying Number of Electrons}

When varying the number of electrons, transport in the source and drain do not significantly change. There still exist regimes with nearly pure sinusoidal motion that do not dephase immediately and regimes with periodic but non-sinusoidal motion that dephase immediately. In addition there still exist minor perturbations in oscillations. Figure 12, which is a system with $12$ sites and $4$ electrons, shows a system in a sinusoidal regime with minor perturbations. 

The phenomena displayed in Figures 10 and 11 still hold true in systems like these, in particular systems with $12$ sites and $4$ electrons. Also, the frequency of the oscillation and the length of the plasmon still vary in the same manner in systems like these, producing the same patterns found in Figures 10 and 11. 

\begin{figure}[h!]
\includegraphics[width=8.6cm]{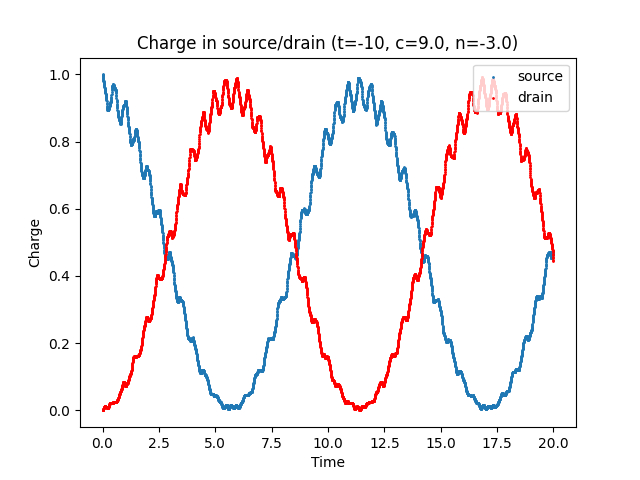}
\caption{ Graphs the transport in the source and drain sites of a system with $t=-10,c=9,n=-3$, $4$ electrons, and $14$ total sites. This system is in a sinusoidal regime and there exist minor perturbations that alter the purely sinusoidal nature of the system. }
\end{figure}

\begin{figure}[h]
\includegraphics[width=8.6cm]{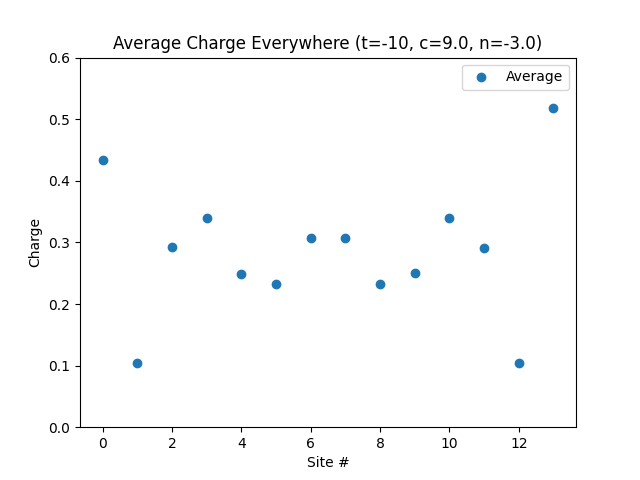}
\caption{Graphs the average charge at each site in a system with 4 electrons, 14 total sites, $t=-10,c=9,n=-3$. There are three local maxima, corresponding to the $4$ electrons. The local maxima decrease the closer they are to the center of the chain. }
\end{figure}

However, the geometric distribution of charge on the chain itself changes, as expected from results before. Figure 13 shows the average geometric charge distribution of the system depicted in Figure 12. There are three local maxima in the chain: about $\frac{1}{3},\frac{1}{2},\frac{2}{3}$ through the chain. These $3$ local maxima correspond to the $4$ electrons in the system, as $4-1=3$, similar to the results found from systems with $12$ sites and $3$ electrons described before. These maxima again occur because the electron-electron interaction is strong enough to prevent all the electrons from filling the middle of the chain. Electrons would want to fill the middle of the chain because they would feel the strongest force from all the positive nuclear sites, but the repulsion from the electron-electron interaction prevents this from happening. Due to the number of electrons, this forms $4-1=3$ local maxima. Moreover, the local maxima in the chain decrease in magnitude the closer they are to the center of the chain. Figure 13 also shows this as the $4$th and $10$th sites, which are the two local maxima furthest from the center of the chain, have a higher magnitude of charge than the $6$th and $7$th sites, which are the local maxima in the center of the chain. This again occurs because of electron-electron interactions. Electrons in the center of the chain feel the strongest electron-electron interaction, so fewer will be found in the center of the chain compared to the edges of the chain. This pattern is found in all systems, including systems with more than $3$ electrons. 

\subsection{Varying Number of Sites}
\begin{figure}[h!]
\includegraphics[width=8.6cm]{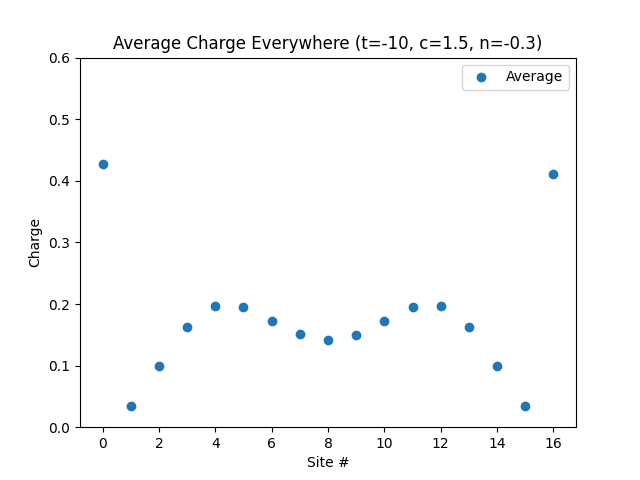}
\caption{Graphs the average charge at each site in a system with $3$ electrons, $17$ total sites, and $t=-10,c=1.5,n=-0.3$. The local maxima and local minima in the chain are decreased when compared to systems with $3$ electrons and $14$ total sites. The average charge in the source and drain also appear slightly decreased. }
\end{figure}

Varying the number of sites, similar to changing the number of electrons, does not produce a significant change in the transport of charge in the source and drain. Similarly, though, there is a noticeable change in the average charge over all time across the chain, but it is less significant in this case than the case where the number of electrons change. Since the number of electrons, $E$, stays constant, the expected number of local maxima should still be $E-1$. But due to the chain being longer or shorter, the expected magnitude of charge at those local maxima should be smaller or larger respectively as charge would optimally distribute along the chain by spreading out as much as possible. This is what is exactly found in results such as those in Figure 14. 

Both the local maxima and local minima are less in magnitude in systems with $17$ total sites, such as the one shown in Figure 14, than in systems with $14$ total sites, such as the one shown in Figure 4. This is caused by the extra length of the chain, allowing electrons to spread out more evenly so the average charge at each site within the chain decreases.

\section{Conclusion}
Through simulating electrons in a one-dimensional chain using the extended Fermi-Hubbard model, plasmons arise with perturbations throughout oscillations due to the one-dimensional geometry of the system. The total charge in the source and drain vary based on the nuclear and Coulomb charges. As the nuclear and Coulomb charges increase, the average net charge in the source and drain decreases while its oscillation reduces in amplitude, and after the midline of the oscillation crosses $1.0$, the oscillation begins increasing in amplitude while the average net charge in the source or drain continues to decrease. Changing the number of electrons and number of sites in the system also change the number of local maxima and the magnitude of charge at these local maxima accordingly. This research both supports prior research done on electron transport using the extended Fermi-Hubbard model while shedding light on undiscovered quantum phenomena \cite{diagram}. 

Future studies should focus on other aspects of systems like these. For instance, observing phase changes, transitions between higher energy states, or two-dimensional systems could provide fruitful results. This research, and other research similar to it, can provide insight into quantum many-body phenomena and further pressing nanotechnology and semiconductor chip research. In fact, systems like these can be thought of as analogs to silicon-doped lattices, with the chain corresponding to phosphorus atoms and the source and drain corresponding to silicon atoms. This is because phosphorus atoms are able to donate an extra electron to generate current in a chain while maintaining the same number of valence electrons silicon has. There are a myriad of real-world applications for this research too. For instance, by understanding charge retention and tunneling in NAND flash memory, higher storage density and longer data retention can be developed \cite{NAND}. In addition, understanding electron flow in silicon-doped lattices can enhance carrier mobility and speed in Silicon-Germanium heterojunction transistors \cite{metal1}. Analyzing electron behavior in doped silicon structures can also enhance plasmonic nanoantenna design and waveguides, developing efficient and controllable quantum light sources for quantum key distribution and quantum information processing \cite{waveguide}.

\begin{acknowledgments}
I wish to acknowledge the support of my mentor, Dr. Garnett Bryant, for advice with the experimental design and analysis of data with this project, and the National Institute for Standards and Technology (NIST) for their assistance with technology to conduct the project. 
\end{acknowledgments}

\medskip
\bibliography{citations}

\begin{thebibliography}{32}%
\makeatletter
\providecommand \@ifxundefined [1]{%
 \@ifx{#1\undefined}
}%
\providecommand \@ifnum [1]{%
 \ifnum #1\expandafter \@firstoftwo
 \else \expandafter \@secondoftwo
 \fi
}%
\providecommand \@ifx [1]{%
 \ifx #1\expandafter \@firstoftwo
 \else \expandafter \@secondoftwo
 \fi
}%
\providecommand \natexlab [1]{#1}%
\providecommand \enquote  [1]{``#1''}%
\providecommand \bibnamefont  [1]{#1}%
\providecommand \bibfnamefont [1]{#1}%
\providecommand \citenamefont [1]{#1}%
\providecommand \href@noop [0]{\@secondoftwo}%
\providecommand \href [0]{\begingroup \@sanitize@url \@href}%
\providecommand \@href[1]{\@@startlink{#1}\@@href}%
\providecommand \@@href[1]{\endgroup#1\@@endlink}%
\providecommand \@sanitize@url [0]{\catcode `\\12\catcode `\$12\catcode `\&12\catcode `\#12\catcode `\^12\catcode `\_12\catcode `\%12\relax}%
\providecommand \@@startlink[1]{}%
\providecommand \@@endlink[0]{}%
\providecommand \url  [0]{\begingroup\@sanitize@url \@url }%
\providecommand \@url [1]{\endgroup\@href {#1}{\urlprefix }}%
\providecommand \urlprefix  [0]{URL }%
\providecommand \Eprint [0]{\href }%
\providecommand \doibase [0]{https://doi.org/}%
\providecommand \selectlanguage [0]{\@gobble}%
\providecommand \bibinfo  [0]{\@secondoftwo}%
\providecommand \bibfield  [0]{\@secondoftwo}%
\providecommand \translation [1]{[#1]}%
\providecommand \BibitemOpen [0]{}%
\providecommand \bibitemStop [0]{}%
\providecommand \bibitemNoStop [0]{.\EOS\space}%
\providecommand \EOS [0]{\spacefactor3000\relax}%
\providecommand \BibitemShut  [1]{\csname bibitem#1\endcsname}%
\let\auto@bib@innerbib\@empty
\bibitem [{\citenamefont {Hirsch}(1989)}]{high-temperature-superconductivity}%
  \BibitemOpen
  \bibfield  {author} {\bibinfo {author} {\bibfnamefont {J.~E.}\ \bibnamefont {Hirsch}},\ }\bibfield  {title} {\bibinfo {title} {Metallic ferromagnetism in a single-band model},\ }\href {https://doi.org/10.1103/PhysRevB.40.2354} {\bibfield  {journal} {\bibinfo  {journal} {Phys. Rev. B}\ }\textbf {\bibinfo {volume} {40}},\ \bibinfo {pages} {2354} (\bibinfo {year} {1989})}\BibitemShut {NoStop}%
\bibitem [{\citenamefont {Imada}\ \emph {et~al.}(1998)\citenamefont {Imada}, \citenamefont {Fujimori},\ and\ \citenamefont {Tokura}}]{mott-insulator}%
  \BibitemOpen
  \bibfield  {author} {\bibinfo {author} {\bibfnamefont {M.}~\bibnamefont {Imada}}, \bibinfo {author} {\bibfnamefont {A.}~\bibnamefont {Fujimori}},\ and\ \bibinfo {author} {\bibfnamefont {Y.}~\bibnamefont {Tokura}},\ }\bibfield  {title} {\bibinfo {title} {Metal-insulator transitions},\ }\href {https://doi.org/10.1103/RevModPhys.70.1039} {\bibfield  {journal} {\bibinfo  {journal} {Rev. Mod. Phys.}\ }\textbf {\bibinfo {volume} {70}},\ \bibinfo {pages} {1039} (\bibinfo {year} {1998})}\BibitemShut {NoStop}%
\bibitem [{\citenamefont {Hasan}\ and\ \citenamefont {Kane}(2010)}]{topo1}%
  \BibitemOpen
  \bibfield  {author} {\bibinfo {author} {\bibfnamefont {M.~Z.}\ \bibnamefont {Hasan}}\ and\ \bibinfo {author} {\bibfnamefont {C.~L.}\ \bibnamefont {Kane}},\ }\bibfield  {title} {\bibinfo {title} {Colloquium: Topological insulators},\ }\href {https://doi.org/10.1103/RevModPhys.82.3045} {\bibfield  {journal} {\bibinfo  {journal} {Rev. Mod. Phys.}\ }\textbf {\bibinfo {volume} {82}},\ \bibinfo {pages} {3045} (\bibinfo {year} {2010})}\BibitemShut {NoStop}%
\bibitem [{\citenamefont {Qi}\ and\ \citenamefont {Zhang}(2011)}]{topo2}%
  \BibitemOpen
  \bibfield  {author} {\bibinfo {author} {\bibfnamefont {X.-L.}\ \bibnamefont {Qi}}\ and\ \bibinfo {author} {\bibfnamefont {S.-C.}\ \bibnamefont {Zhang}},\ }\bibfield  {title} {\bibinfo {title} {Topological insulators and superconductors},\ }\href {https://doi.org/10.1103/RevModPhys.83.1057} {\bibfield  {journal} {\bibinfo  {journal} {Rev. Mod. Phys.}\ }\textbf {\bibinfo {volume} {83}},\ \bibinfo {pages} {1057} (\bibinfo {year} {2011})}\BibitemShut {NoStop}%
\bibitem [{\citenamefont {Jotzu}\ \emph {et~al.}(2014)\citenamefont {Jotzu}, \citenamefont {Messer}, \citenamefont {Desbuquois}, \citenamefont {Lebrat}, \citenamefont {Uehlinger}, \citenamefont {Greif},\ and\ \citenamefont {Esslinger}}]{topo3}%
  \BibitemOpen
  \bibfield  {author} {\bibinfo {author} {\bibfnamefont {G.}~\bibnamefont {Jotzu}}, \bibinfo {author} {\bibfnamefont {M.}~\bibnamefont {Messer}}, \bibinfo {author} {\bibfnamefont {R.}~\bibnamefont {Desbuquois}}, \bibinfo {author} {\bibfnamefont {M.}~\bibnamefont {Lebrat}}, \bibinfo {author} {\bibfnamefont {T.}~\bibnamefont {Uehlinger}}, \bibinfo {author} {\bibfnamefont {D.}~\bibnamefont {Greif}},\ and\ \bibinfo {author} {\bibfnamefont {T.}~\bibnamefont {Esslinger}},\ }\bibfield  {title} {\bibinfo {title} {Experimental realization of the topological haldane model with ultracold fermions},\ }\href {https://doi.org/10.1038/nature13915} {\bibfield  {journal} {\bibinfo  {journal} {Nature}\ }\textbf {\bibinfo {volume} {515}},\ \bibinfo {pages} {237} (\bibinfo {year} {2014})}\BibitemShut {NoStop}%
\bibitem [{\citenamefont {Wang}\ \emph {et~al.}(2018)\citenamefont {Wang}, \citenamefont {Zhang}, \citenamefont {Hu},\ and\ \citenamefont {Shi}}]{transistor}%
  \BibitemOpen
  \bibfield  {author} {\bibinfo {author} {\bibfnamefont {B.}~\bibnamefont {Wang}}, \bibinfo {author} {\bibfnamefont {H.-M.}\ \bibnamefont {Zhang}}, \bibinfo {author} {\bibfnamefont {H.-Y.}\ \bibnamefont {Hu}},\ and\ \bibinfo {author} {\bibfnamefont {X.-W.}\ \bibnamefont {Shi}},\ }\bibfield  {title} {\bibinfo {title} {Enhancement of off-state characteristics in junctionless field effect transistor using a field plate},\ }\href {https://doi.org/10.1088/1674-1056/27/6/067402} {\bibfield  {journal} {\bibinfo  {journal} {Chinese Physics B}\ }\textbf {\bibinfo {volume} {27}},\ \bibinfo {eid} {067402} (\bibinfo {year} {2018})}\BibitemShut {NoStop}%
\bibitem [{\citenamefont {Nilius}\ \emph {et~al.}(2002)\citenamefont {Nilius}, \citenamefont {Wallis},\ and\ \citenamefont {Ho}}]{metal1}%
  \BibitemOpen
  \bibfield  {author} {\bibinfo {author} {\bibfnamefont {N.}~\bibnamefont {Nilius}}, \bibinfo {author} {\bibfnamefont {T.~M.}\ \bibnamefont {Wallis}},\ and\ \bibinfo {author} {\bibfnamefont {W.}~\bibnamefont {Ho}},\ }\bibfield  {title} {\bibinfo {title} {Development of one-dimensional band structure in artificial gold chains},\ }\href {https://doi.org/10.1126/science.1075242} {\bibfield  {journal} {\bibinfo  {journal} {Science}\ }\textbf {\bibinfo {volume} {297}},\ \bibinfo {pages} {1853} (\bibinfo {year} {2002})},\ \Eprint {https://arxiv.org/abs/https://www.science.org/doi/pdf/10.1126/science.1075242} {https://www.science.org/doi/pdf/10.1126/science.1075242} \BibitemShut {NoStop}%
\bibitem [{\citenamefont {Nazin}\ \emph {et~al.}(2003)\citenamefont {Nazin}, \citenamefont {Qiu},\ and\ \citenamefont {Ho}}]{metal2}%
  \BibitemOpen
  \bibfield  {author} {\bibinfo {author} {\bibfnamefont {G.~V.}\ \bibnamefont {Nazin}}, \bibinfo {author} {\bibfnamefont {X.~H.}\ \bibnamefont {Qiu}},\ and\ \bibinfo {author} {\bibfnamefont {W.}~\bibnamefont {Ho}},\ }\bibfield  {title} {\bibinfo {title} {Visualization and spectroscopy of a metal-molecule-metal bridge},\ }\href {https://doi.org/10.1126/science.1088971} {\bibfield  {journal} {\bibinfo  {journal} {Science}\ }\textbf {\bibinfo {volume} {302}},\ \bibinfo {pages} {77} (\bibinfo {year} {2003})},\ \Eprint {https://arxiv.org/abs/https://www.science.org/doi/pdf/10.1126/science.1088971} {https://www.science.org/doi/pdf/10.1126/science.1088971} \BibitemShut {NoStop}%
\bibitem [{\citenamefont {F\"olsch}\ \emph {et~al.}(2004)\citenamefont {F\"olsch}, \citenamefont {Hyldgaard}, \citenamefont {Koch},\ and\ \citenamefont {Ploog}}]{metal3}%
  \BibitemOpen
  \bibfield  {author} {\bibinfo {author} {\bibfnamefont {S.}~\bibnamefont {F\"olsch}}, \bibinfo {author} {\bibfnamefont {P.}~\bibnamefont {Hyldgaard}}, \bibinfo {author} {\bibfnamefont {R.}~\bibnamefont {Koch}},\ and\ \bibinfo {author} {\bibfnamefont {K.~H.}\ \bibnamefont {Ploog}},\ }\bibfield  {title} {\bibinfo {title} {Quantum confinement in monatomic cu chains on cu(111)},\ }\href {https://doi.org/10.1103/PhysRevLett.92.056803} {\bibfield  {journal} {\bibinfo  {journal} {Phys. Rev. Lett.}\ }\textbf {\bibinfo {volume} {92}},\ \bibinfo {pages} {056803} (\bibinfo {year} {2004})}\BibitemShut {NoStop}%
\bibitem [{\citenamefont {Salfi}\ \emph {et~al.}(2016)\citenamefont {Salfi}, \citenamefont {Mol}, \citenamefont {Rahman}, \citenamefont {Klimeck}, \citenamefont {Simmons}, \citenamefont {Hollenberg},\ and\ \citenamefont {Rogge}}]{silicon1}%
  \BibitemOpen
  \bibfield  {author} {\bibinfo {author} {\bibfnamefont {J.}~\bibnamefont {Salfi}}, \bibinfo {author} {\bibfnamefont {J.~A.}\ \bibnamefont {Mol}}, \bibinfo {author} {\bibfnamefont {R.}~\bibnamefont {Rahman}}, \bibinfo {author} {\bibfnamefont {G.}~\bibnamefont {Klimeck}}, \bibinfo {author} {\bibfnamefont {M.~Y.}\ \bibnamefont {Simmons}}, \bibinfo {author} {\bibfnamefont {L.~C.~L.}\ \bibnamefont {Hollenberg}},\ and\ \bibinfo {author} {\bibfnamefont {S.}~\bibnamefont {Rogge}},\ }\bibfield  {title} {\bibinfo {title} {Quantum simulation of the hubbard model with dopant atoms in silicon},\ }\bibfield  {journal} {\bibinfo  {journal} {Nature Communications}\ }\textbf {\bibinfo {volume} {7}},\ \href {https://doi.org/10.1038/ncomms11342} {10.1038/ncomms11342} (\bibinfo {year} {2016})\BibitemShut {NoStop}%
\bibitem [{\citenamefont {Dusko}\ \emph {et~al.}(2018)\citenamefont {Dusko}, \citenamefont {Delgado}, \citenamefont {Saraiva},\ and\ \citenamefont {Koiller}}]{silicon2}%
  \BibitemOpen
  \bibfield  {author} {\bibinfo {author} {\bibfnamefont {A.}~\bibnamefont {Dusko}}, \bibinfo {author} {\bibfnamefont {A.}~\bibnamefont {Delgado}}, \bibinfo {author} {\bibfnamefont {A.}~\bibnamefont {Saraiva}},\ and\ \bibinfo {author} {\bibfnamefont {B.}~\bibnamefont {Koiller}},\ }\bibfield  {title} {\bibinfo {title} {Adequacy of si:p chains as fermi{\textendash}hubbard simulators},\ }\bibfield  {journal} {\bibinfo  {journal} {npj Quantum Information}\ }\textbf {\bibinfo {volume} {4}},\ \href {https://doi.org/10.1038/s41534-017-0051-1} {10.1038/s41534-017-0051-1} (\bibinfo {year} {2018})\BibitemShut {NoStop}%
\bibitem [{\citenamefont {Le}\ \emph {et~al.}(2017)\citenamefont {Le}, \citenamefont {Fisher},\ and\ \citenamefont {Ginossar}}]{silicon3}%
  \BibitemOpen
  \bibfield  {author} {\bibinfo {author} {\bibfnamefont {N.~H.}\ \bibnamefont {Le}}, \bibinfo {author} {\bibfnamefont {A.~J.}\ \bibnamefont {Fisher}},\ and\ \bibinfo {author} {\bibfnamefont {E.}~\bibnamefont {Ginossar}},\ }\bibfield  {title} {\bibinfo {title} {Extended hubbard model for mesoscopic transport in donor arrays in silicon},\ }\href {https://doi.org/10.1103/PhysRevB.96.245406} {\bibfield  {journal} {\bibinfo  {journal} {Phys. Rev. B}\ }\textbf {\bibinfo {volume} {96}},\ \bibinfo {pages} {245406} (\bibinfo {year} {2017})}\BibitemShut {NoStop}%
\bibitem [{\citenamefont {Sigillito}\ \emph {et~al.}(2019)\citenamefont {Sigillito}, \citenamefont {Loy}, \citenamefont {Zajac}, \citenamefont {Gullans}, \citenamefont {Edge},\ and\ \citenamefont {Petta}}]{dot1}%
  \BibitemOpen
  \bibfield  {author} {\bibinfo {author} {\bibfnamefont {A.}~\bibnamefont {Sigillito}}, \bibinfo {author} {\bibfnamefont {J.}~\bibnamefont {Loy}}, \bibinfo {author} {\bibfnamefont {D.}~\bibnamefont {Zajac}}, \bibinfo {author} {\bibfnamefont {M.}~\bibnamefont {Gullans}}, \bibinfo {author} {\bibfnamefont {L.}~\bibnamefont {Edge}},\ and\ \bibinfo {author} {\bibfnamefont {J.}~\bibnamefont {Petta}},\ }\bibfield  {title} {\bibinfo {title} {Site-selective quantum control in an isotopically enriched $^{28}\mathrm{Si}/{\mathrm{si}}_{0.7}{\mathrm{ge}}_{0.3}$ quadruple quantum dot},\ }\href {https://doi.org/10.1103/PhysRevApplied.11.061006} {\bibfield  {journal} {\bibinfo  {journal} {Phys. Rev. Appl.}\ }\textbf {\bibinfo {volume} {11}},\ \bibinfo {pages} {061006} (\bibinfo {year} {2019})}\BibitemShut {NoStop}%
\bibitem [{\citenamefont {Zajac}\ \emph {et~al.}(2016)\citenamefont {Zajac}, \citenamefont {Hazard}, \citenamefont {Mi}, \citenamefont {Nielsen},\ and\ \citenamefont {Petta}}]{dot2}%
  \BibitemOpen
  \bibfield  {author} {\bibinfo {author} {\bibfnamefont {D.~M.}\ \bibnamefont {Zajac}}, \bibinfo {author} {\bibfnamefont {T.~M.}\ \bibnamefont {Hazard}}, \bibinfo {author} {\bibfnamefont {X.}~\bibnamefont {Mi}}, \bibinfo {author} {\bibfnamefont {E.}~\bibnamefont {Nielsen}},\ and\ \bibinfo {author} {\bibfnamefont {J.~R.}\ \bibnamefont {Petta}},\ }\bibfield  {title} {\bibinfo {title} {Scalable gate architecture for a one-dimensional array of semiconductor spin qubits},\ }\href {https://doi.org/10.1103/PhysRevApplied.6.054013} {\bibfield  {journal} {\bibinfo  {journal} {Phys. Rev. Appl.}\ }\textbf {\bibinfo {volume} {6}},\ \bibinfo {pages} {054013} (\bibinfo {year} {2016})}\BibitemShut {NoStop}%
\bibitem [{\citenamefont {Haider}\ \emph {et~al.}(2009)\citenamefont {Haider}, \citenamefont {Pitters}, \citenamefont {DiLabio}, \citenamefont {Livadaru}, \citenamefont {Mutus},\ and\ \citenamefont {Wolkow}}]{dangling1}%
  \BibitemOpen
  \bibfield  {author} {\bibinfo {author} {\bibfnamefont {M.~B.}\ \bibnamefont {Haider}}, \bibinfo {author} {\bibfnamefont {J.~L.}\ \bibnamefont {Pitters}}, \bibinfo {author} {\bibfnamefont {G.~A.}\ \bibnamefont {DiLabio}}, \bibinfo {author} {\bibfnamefont {L.}~\bibnamefont {Livadaru}}, \bibinfo {author} {\bibfnamefont {J.~Y.}\ \bibnamefont {Mutus}},\ and\ \bibinfo {author} {\bibfnamefont {R.~A.}\ \bibnamefont {Wolkow}},\ }\bibfield  {title} {\bibinfo {title} {Controlled coupling and occupation of silicon atomic quantum dots at room temperature},\ }\href {https://doi.org/10.1103/PhysRevLett.102.046805} {\bibfield  {journal} {\bibinfo  {journal} {Phys. Rev. Lett.}\ }\textbf {\bibinfo {volume} {102}},\ \bibinfo {pages} {046805} (\bibinfo {year} {2009})}\BibitemShut {NoStop}%
\bibitem [{\citenamefont {Schofield}\ \emph {et~al.}(2013)\citenamefont {Schofield}, \citenamefont {Studer}, \citenamefont {Hirjibehedin}, \citenamefont {Curson}, \citenamefont {Aeppli},\ and\ \citenamefont {Bowler}}]{dangling2}%
  \BibitemOpen
  \bibfield  {author} {\bibinfo {author} {\bibfnamefont {S.~R.}\ \bibnamefont {Schofield}}, \bibinfo {author} {\bibfnamefont {P.}~\bibnamefont {Studer}}, \bibinfo {author} {\bibfnamefont {C.~F.}\ \bibnamefont {Hirjibehedin}}, \bibinfo {author} {\bibfnamefont {N.~J.}\ \bibnamefont {Curson}}, \bibinfo {author} {\bibfnamefont {G.}~\bibnamefont {Aeppli}},\ and\ \bibinfo {author} {\bibfnamefont {D.~R.}\ \bibnamefont {Bowler}},\ }\bibfield  {title} {\bibinfo {title} {Quantum engineering at the silicon surface using dangling bonds},\ }\href {https://doi.org/10.1038/ncomms2679} {\bibfield  {journal} {\bibinfo  {journal} {Nature Communications}\ }\textbf {\bibinfo {volume} {4}},\ \bibinfo {pages} {1649} (\bibinfo {year} {2013})}\BibitemShut {NoStop}%
\bibitem [{\citenamefont {Wyrick}\ \emph {et~al.}(2018)\citenamefont {Wyrick}, \citenamefont {Wang}, \citenamefont {Namboodiri}, \citenamefont {Schmucker}, \citenamefont {Kashid},\ and\ \citenamefont {Silver}}]{dangling3}%
  \BibitemOpen
  \bibfield  {author} {\bibinfo {author} {\bibfnamefont {J.}~\bibnamefont {Wyrick}}, \bibinfo {author} {\bibfnamefont {X.}~\bibnamefont {Wang}}, \bibinfo {author} {\bibfnamefont {P.}~\bibnamefont {Namboodiri}}, \bibinfo {author} {\bibfnamefont {S.}~\bibnamefont {Schmucker}}, \bibinfo {author} {\bibfnamefont {R.}~\bibnamefont {Kashid}},\ and\ \bibinfo {author} {\bibfnamefont {R.}~\bibnamefont {Silver}},\ }\bibfield  {title} {\bibinfo {title} {Atom-by-atom construction of a cyclic artificial molecule in silicon}\ }\href {https://doi.org/https://doi.org/10.1021/acs.nanolett.8b02919} {https://doi.org/10.1021/acs.nanolett.8b02919} (\bibinfo {year} {2018})\BibitemShut {NoStop}%
\bibitem [{\citenamefont {Wang}\ \emph {et~al.}(2022)\citenamefont {Wang}, \citenamefont {Khatami}, \citenamefont {Fei}, \citenamefont {Wyrick}, \citenamefont {Namboodiri}, \citenamefont {Kashid}, \citenamefont {Rigosi}, \citenamefont {Bryant},\ and\ \citenamefont {Silver}}]{silicon4}%
  \BibitemOpen
  \bibfield  {author} {\bibinfo {author} {\bibfnamefont {X.}~\bibnamefont {Wang}}, \bibinfo {author} {\bibfnamefont {E.}~\bibnamefont {Khatami}}, \bibinfo {author} {\bibfnamefont {F.}~\bibnamefont {Fei}}, \bibinfo {author} {\bibfnamefont {J.}~\bibnamefont {Wyrick}}, \bibinfo {author} {\bibfnamefont {P.}~\bibnamefont {Namboodiri}}, \bibinfo {author} {\bibfnamefont {R.}~\bibnamefont {Kashid}}, \bibinfo {author} {\bibfnamefont {A.~F.}\ \bibnamefont {Rigosi}}, \bibinfo {author} {\bibfnamefont {G.}~\bibnamefont {Bryant}},\ and\ \bibinfo {author} {\bibfnamefont {R.}~\bibnamefont {Silver}},\ }\bibfield  {title} {\bibinfo {title} {Experimental realization of an extended fermi-hubbard model using a 2d lattice of dopant-based quantum dots},\ }\bibfield  {journal} {\bibinfo  {journal} {Nature Communications}\ }\textbf {\bibinfo {volume} {13}},\ \href {https://doi.org/10.1038/s41467-022-34220-w} {10.1038/s41467-022-34220-w} (\bibinfo {year} {2022})\BibitemShut {NoStop}%
\bibitem [{hub(1963)}]{hubbard1}%
  \BibitemOpen
  \bibfield  {title} {\bibinfo {title} {Electron correlations in narrow energy bands},\ }\href {https://doi.org/10.1098/rspa.1963.0204} {\bibfield  {journal} {\bibinfo  {journal} {Proceedings of the Royal Society of London. Series A. Mathematical and Physical Sciences}\ }\textbf {\bibinfo {volume} {276}},\ \bibinfo {pages} {238} (\bibinfo {year} {1963})}\BibitemShut {NoStop}%
\bibitem [{\citenamefont {Gutzwiller}(1963)}]{hubbard2}%
  \BibitemOpen
  \bibfield  {author} {\bibinfo {author} {\bibfnamefont {M.~C.}\ \bibnamefont {Gutzwiller}},\ }\bibfield  {title} {\bibinfo {title} {Effect of correlation on the ferromagnetism of transition metals},\ }\href {https://doi.org/10.1103/PhysRevLett.10.159} {\bibfield  {journal} {\bibinfo  {journal} {Phys. Rev. Lett.}\ }\textbf {\bibinfo {volume} {10}},\ \bibinfo {pages} {159} (\bibinfo {year} {1963})}\BibitemShut {NoStop}%
\bibitem [{hub(1964)}]{hubbard3}%
  \BibitemOpen
  \bibfield  {title} {\bibinfo {title} {Electron correlations in narrow energy bands. {II}. the degenerate band case},\ }\href {https://doi.org/10.1098/rspa.1964.0019} {\bibfield  {journal} {\bibinfo  {journal} {Proceedings of the Royal Society of London. Series A. Mathematical and Physical Sciences}\ }\textbf {\bibinfo {volume} {277}},\ \bibinfo {pages} {237} (\bibinfo {year} {1964})}\BibitemShut {NoStop}%
\bibitem [{\citenamefont {Gull}\ \emph {et~al.}(2013)\citenamefont {Gull}, \citenamefont {Parcollet},\ and\ \citenamefont {Millis}}]{unconventionalsuper}%
  \BibitemOpen
  \bibfield  {author} {\bibinfo {author} {\bibfnamefont {E.}~\bibnamefont {Gull}}, \bibinfo {author} {\bibfnamefont {O.}~\bibnamefont {Parcollet}},\ and\ \bibinfo {author} {\bibfnamefont {A.~J.}\ \bibnamefont {Millis}},\ }\bibfield  {title} {\bibinfo {title} {Superconductivity and the pseudogap in the two-dimensional hubbard model},\ }\href {https://doi.org/10.1103/PhysRevLett.110.216405} {\bibfield  {journal} {\bibinfo  {journal} {Phys. Rev. Lett.}\ }\textbf {\bibinfo {volume} {110}},\ \bibinfo {pages} {216405} (\bibinfo {year} {2013})}\BibitemShut {NoStop}%
\bibitem [{\citenamefont {Balents}(2010)}]{spinliquid}%
  \BibitemOpen
  \bibfield  {author} {\bibinfo {author} {\bibfnamefont {L.}~\bibnamefont {Balents}},\ }\bibfield  {title} {\bibinfo {title} {Spin liquids in frustrated magnets},\ }\href {https://doi.org/10.1038/nature08917} {\bibfield  {journal} {\bibinfo  {journal} {Nature}\ }\textbf {\bibinfo {volume} {464}},\ \bibinfo {pages} {199} (\bibinfo {year} {2010})}\BibitemShut {NoStop}%
\bibitem [{\citenamefont {Tasaki}(1998)}]{Nferromagnetism}%
  \BibitemOpen
  \bibfield  {author} {\bibinfo {author} {\bibfnamefont {H.}~\bibnamefont {Tasaki}},\ }\bibfield  {title} {\bibinfo {title} {From nagaoka{\textquotesingle}s ferromagnetism to flat-band ferromagnetism and beyond: An introduction to ferromagnetism in the hubbard model},\ }\href {https://doi.org/10.1143/ptp.99.489} {\bibfield  {journal} {\bibinfo  {journal} {Progress of Theoretical Physics}\ }\textbf {\bibinfo {volume} {99}},\ \bibinfo {pages} {489} (\bibinfo {year} {1998})}\BibitemShut {NoStop}%
\bibitem [{\citenamefont {Dutta}\ \emph {et~al.}(2013)\citenamefont {Dutta}, \citenamefont {Sowi\ifmmode~\acute{n}\else \'{n}\fi{}ski},\ and\ \citenamefont {Lewenstein}}]{coldfermion1}%
  \BibitemOpen
  \bibfield  {author} {\bibinfo {author} {\bibfnamefont {O.}~\bibnamefont {Dutta}}, \bibinfo {author} {\bibfnamefont {T.}~\bibnamefont {Sowi\ifmmode~\acute{n}\else \'{n}\fi{}ski}},\ and\ \bibinfo {author} {\bibfnamefont {M.}~\bibnamefont {Lewenstein}},\ }\bibfield  {title} {\bibinfo {title} {Orbital physics of polar fermi molecules},\ }\href {https://doi.org/10.1103/PhysRevA.87.023619} {\bibfield  {journal} {\bibinfo  {journal} {Phys. Rev. A}\ }\textbf {\bibinfo {volume} {87}},\ \bibinfo {pages} {023619} (\bibinfo {year} {2013})}\BibitemShut {NoStop}%
\bibitem [{\citenamefont {Dutta}\ \emph {et~al.}(2015)\citenamefont {Dutta}, \citenamefont {Gajda}, \citenamefont {Hauke}, \citenamefont {Lewenstein}, \citenamefont {L\"{u}hmann}, \citenamefont {Malomed}, \citenamefont {Sowi{\'{n}}ski},\ and\ \citenamefont {Zakrzewski}}]{coldfermion2}%
  \BibitemOpen
  \bibfield  {author} {\bibinfo {author} {\bibfnamefont {O.}~\bibnamefont {Dutta}}, \bibinfo {author} {\bibfnamefont {M.}~\bibnamefont {Gajda}}, \bibinfo {author} {\bibfnamefont {P.}~\bibnamefont {Hauke}}, \bibinfo {author} {\bibfnamefont {M.}~\bibnamefont {Lewenstein}}, \bibinfo {author} {\bibfnamefont {D.-S.}\ \bibnamefont {L\"{u}hmann}}, \bibinfo {author} {\bibfnamefont {B.~A.}\ \bibnamefont {Malomed}}, \bibinfo {author} {\bibfnamefont {T.}~\bibnamefont {Sowi{\'{n}}ski}},\ and\ \bibinfo {author} {\bibfnamefont {J.}~\bibnamefont {Zakrzewski}},\ }\bibfield  {title} {\bibinfo {title} {Non-standard hubbard models in optical lattices: a review},\ }\href {https://doi.org/10.1088/0034-4885/78/6/066001} {\bibfield  {journal} {\bibinfo  {journal} {Reports on Progress in Physics}\ }\textbf {\bibinfo {volume} {78}},\ \bibinfo {pages} {066001} (\bibinfo {year} {2015})}\BibitemShut {NoStop}%
\bibitem [{\citenamefont {van Loon}\ \emph {et~al.}(2015)\citenamefont {van Loon}, \citenamefont {Katsnelson},\ and\ \citenamefont {Lemeshko}}]{coldfermion3}%
  \BibitemOpen
  \bibfield  {author} {\bibinfo {author} {\bibfnamefont {E.~G. C.~P.}\ \bibnamefont {van Loon}}, \bibinfo {author} {\bibfnamefont {M.~I.}\ \bibnamefont {Katsnelson}},\ and\ \bibinfo {author} {\bibfnamefont {M.}~\bibnamefont {Lemeshko}},\ }\bibfield  {title} {\bibinfo {title} {Ultralong-range order in the fermi-hubbard model with long-range interactions},\ }\href {https://doi.org/10.1103/PhysRevB.92.081106} {\bibfield  {journal} {\bibinfo  {journal} {Phys. Rev. B}\ }\textbf {\bibinfo {volume} {92}},\ \bibinfo {pages} {081106} (\bibinfo {year} {2015})}\BibitemShut {NoStop}%
\bibitem [{\citenamefont {van Loon}\ \emph {et~al.}(2018)\citenamefont {van Loon}, \citenamefont {Rösner}, \citenamefont {Schönhoff}, \citenamefont {Katsnelson},\ and\ \citenamefont {Wehling}}]{transitionmetaldi}%
  \BibitemOpen
  \bibfield  {author} {\bibinfo {author} {\bibfnamefont {E.~G. C.~P.}\ \bibnamefont {van Loon}}, \bibinfo {author} {\bibfnamefont {M.}~\bibnamefont {Rösner}}, \bibinfo {author} {\bibfnamefont {G.}~\bibnamefont {Schönhoff}}, \bibinfo {author} {\bibfnamefont {M.~I.}\ \bibnamefont {Katsnelson}},\ and\ \bibinfo {author} {\bibfnamefont {T.~O.}\ \bibnamefont {Wehling}},\ }\bibfield  {title} {\bibinfo {title} {Competing coulomb and electron{\textendash}phonon interactions in {NbS}2},\ }\bibfield  {journal} {\bibinfo  {journal} {npj Quantum Materials}\ }\textbf {\bibinfo {volume} {3}},\ \href {https://doi.org/10.1038/s41535-018-0105-4} {10.1038/s41535-018-0105-4} (\bibinfo {year} {2018})\BibitemShut {NoStop}%
\bibitem [{\citenamefont {Bryant}\ \emph {et~al.}()\citenamefont {Bryant}, \citenamefont {Townsend}, \citenamefont {Neuman}, \citenamefont {Debrecht},\ and\ \citenamefont {Aizpurua}}]{diagram}%
  \BibitemOpen
  \bibfield  {author} {\bibinfo {author} {\bibfnamefont {G.}~\bibnamefont {Bryant}}, \bibinfo {author} {\bibfnamefont {E.}~\bibnamefont {Townsend}}, \bibinfo {author} {\bibfnamefont {T.}~\bibnamefont {Neuman}}, \bibinfo {author} {\bibfnamefont {A.}~\bibnamefont {Debrecht}},\ and\ \bibinfo {author} {\bibfnamefont {J.}~\bibnamefont {Aizpurua}},\ }\bibfield  {title} {\bibinfo {title} {The emergence of atomic-scale quantum plasmonics},\ }\bibinfo {note} {preprint from July 2020}\BibitemShut {NoStop}%
\bibitem [{\citenamefont {El-Mikkawy}(2004)}]{determinant}%
  \BibitemOpen
  \bibfield  {author} {\bibinfo {author} {\bibfnamefont {M.~E.}\ \bibnamefont {El-Mikkawy}},\ }\bibfield  {title} {\bibinfo {title} {On the inverse of a general tridiagonal matrix},\ }\href {https://doi.org/10.1016/s0096-3003(03)00298-4} {\bibfield  {journal} {\bibinfo  {journal} {Applied Mathematics and Computation}\ }\textbf {\bibinfo {volume} {150}},\ \bibinfo {pages} {669–679} (\bibinfo {year} {2004})}\BibitemShut {NoStop}%
\bibitem [{\citenamefont {Stoddart}(2018)}]{NAND}%
  \BibitemOpen
  \bibfield  {author} {\bibinfo {author} {\bibfnamefont {A.}~\bibnamefont {Stoddart}},\ }\bibfield  {title} {\bibinfo {title} {Electronic devices: Finding flaws in a flash},\ }\bibfield  {journal} {\bibinfo  {journal} {Nature Reviews Materials}\ }\textbf {\bibinfo {volume} {3}},\ \href {https://doi.org/10.1038/natrevmats.2018.1} {10.1038/natrevmats.2018.1} (\bibinfo {year} {2018})\BibitemShut {NoStop}%
\bibitem [{\citenamefont {Khodadadi}\ \emph {et~al.}(2020)\citenamefont {Khodadadi}, \citenamefont {Nozhat},\ and\ \citenamefont {Moshiri}}]{waveguide}%
  \BibitemOpen
  \bibfield  {author} {\bibinfo {author} {\bibfnamefont {M.}~\bibnamefont {Khodadadi}}, \bibinfo {author} {\bibfnamefont {N.}~\bibnamefont {Nozhat}},\ and\ \bibinfo {author} {\bibfnamefont {S.~M.~M.}\ \bibnamefont {Moshiri}},\ }\bibfield  {title} {\bibinfo {title} {Analytic approach to study a hybrid plasmonic waveguide-fed and numerically design a nano-antenna based on the new director},\ }\href {https://doi.org/10.1364/OE.373221} {\bibfield  {journal} {\bibinfo  {journal} {Opt. Express}\ }\textbf {\bibinfo {volume} {28}},\ \bibinfo {pages} {3305} (\bibinfo {year} {2020})}\BibitemShut {NoStop}%
\end{thebibliography}%
\end{document}